\documentclass[journal, final, 9pt, twocolumn]{IEEEtran} 

\usepackage[nice]{nicefrac}
\usepackage{graphicx}
\usepackage{psfrag}
\usepackage{dcolumn}
\usepackage{epsf}
\usepackage{float}
\usepackage{amssymb}
\usepackage{amsmath}
\usepackage{amsfonts}
\usepackage{cite}
\usepackage[tight,footnotesize]{subfigure} 
\usepackage{dblfloatfix}
\usepackage{acronym}
\usepackage[usenames,dvipsnames]{color}
\usepackage{multirow}
\usepackage{ctable}
\usepackage{float}
\usepackage{placeins}
\usepackage{acronym}
\usepackage{bm}
\usepackage{caption}
\usepackage{psfrag}
\usepackage{booktabs, balance}

\usepackage{algpseudocode}
\usepackage{algorithmicx}
\usepackage{algorithm}

\usepackage{setspace}

\newlength{\dhatheight}

\usepackage[normalem]{ulem}

\definecolor{mygreen}{rgb}{0.2,0.6,0.2}

\definecolor{myred}{rgb}{0.8,0.2,0.2}

\definecolor{mygreen}{rgb}{0.2,0.8,0.2}

\renewcommand{\bm}[1]{{\mathbf #1}}

\definecolor{myred}{rgb}{0.9,0.1,0.1}

\begin{document}
	
\title{\huge \bf Multi-Speaker DOA Estimation Using Deep Convolutional Networks Trained with Noise Signals}

\author{{Soumitro Chakrabarty, \emph{Student Member, IEEE}, and Emanu\"{e}l A.~P.~Habets, \emph{Senior Member, IEEE}}
	
	\thanks{Copyright (c) 2018 IEEE. Personal use of this material is permitted. However, permission to use this material for any other purposes must be obtained from the IEEE by sending a request to pubs-permissions@ieee.org.}
	\thanks{Soumitro Chakrabarty and Emanu\"{e}l Habets are with the International Audio Laboratories Erlangen (a joint institution between the University of Erlangen-Nuremberg and Fraunhofer IIS), Germany.}
	\thanks{Corresponding address: Soumitro Chakrabarty, International Audio Laboratories Erlangen, University of Erlangen-Nuremberg, Am Wolfsmantel 33, 91058 Erlangen, Germany. Email: soumitro.chakrabarty@audiolabs-erlangen.de.}
	\thanks{This manuscript has been submitted to IEEE for possible publication. Based on the policies of IEEE, following a Copyright transfer, this version of the article may not be available.}
} \markboth{}{Chakrabarty and Habets, CNN based mulit-speaker localization}

\setlength{\belowdisplayskip}{1mm}
\setlength{\belowcaptionskip}{-1mm} 

\captionsetup{width=8cm}

\maketitle

\begin{abstract}
Supervised learning based methods for source localization, being data driven, can be adapted to different acoustic conditions via training and have been shown to be robust to adverse acoustic environments. In this paper, a convolutional neural network (CNN) based supervised learning method for estimating the direction-of-arrival (DOA) of multiple speakers is proposed. Multi-speaker DOA estimation is formulated as a multi-class multi-label classification problem, where the assignment of each DOA label to the input feature is treated as a separate binary classification problem. The phase component of the short-time Fourier transform (STFT) coefficients of the received microphone signals are directly fed into the CNN, and the features for DOA estimation are learnt during training. Utilizing the assumption of disjoint speaker activity in the STFT domain, a novel method is proposed to train the CNN with synthesized noise signals. Through experimental evaluation with both simulated and measured acoustic impulse responses, the ability of the proposed DOA estimation approach to adapt to unseen acoustic conditions and its robustness to unseen noise type is demonstrated. Through additional empirical investigation, it is also shown that with an array of M microphones our proposed framework yields the best localization performance with M-1 convolution layers. The ability of the proposed method to accurately localize speakers in a dynamic acoustic scenario with varying number of sources is also shown.  
\end{abstract}


\begin{sloppy}

\section{Introduction}
\label{sec:Intro}

Many applications such as hands-free communication, teleconferencing, robot audition and distant speech recognition require information on the location of sound sources in the acoustic environment. Information regarding the source location can be utilized for the task of enhancing the signal coming from a specific location while suppressing the undesired signal components. In some applications, the information is used for camera steering whereas in applications like robot audition the source location information is used for navigation purposes. The relative direction of a sound source with respect to a microphone array is generally given in terms of the direction of arrival (DOA) of the sound wave originating from the source position.  In most practical scenarios, this information is not available and the DOA of the sound source need to be estimated. However, accurate DOA estimation is a challenging task in the presence of noise and reverberation. The task becomes even more difficult when the DOAs of multiple sound sources need to be estimated. 

In the literature related to DOA estimation, there exist two kinds of estimation paradigms: broadband and narrowband DOA estimation. In narrowband DOA estimation, the task of DOA estimation is performed separately for each frequency sub-band, whereas in broadband DOA estimation the task is performed for the whole input spectrum. In this work, the focus is on broadband DOA estimation.  

Over the years, several approaches have been developed for the task of broadband DOA estimation. Some popular approaches are: \emph{i)} subspace based approaches such as multiple signal classification (MUSIC) \cite{Schmidt1986,Dmochowski2007}, \emph{ii)} time difference of arrival (TDOA) based approaches that use the family of generalized cross correlation (GCC) methods \cite{Knapp1976, Huang2001b}, \emph{iii)} generalizations of the cross-correlation methods such as steered response power with phase transform (SRP-PHAT) \cite{Brandstein1997}, and multichannel cross correlation coefficient (MCCC) \cite{Benesty2008a}, \emph{iv)} adaptive multichannel time delay estimation using blind system identification based methods \cite{Benesty2003a}, \emph{v)} probabilistic model based methods such as maximum likelihood method \cite{Stoica1990} and \emph{vi)} methods based on histogram analysis of narrowband DOA estimates \cite{Delikaris-Manias2017, Moore2016}. These methods are generally formulated under the assumption of free-field propagation of sound waves, however in indoor acoustic environments this assumption is violated due to the presence of reverberation leading to severe degradation in their performance. Additionally, these methods are also not robust to noise and generally have a high computational cost \cite{Benesty2008a}. 

Compared to the signal processing based approaches, supervised learning approaches, being data driven, have the advantage that they can be adapted to different acoustic conditions via training. Also, if training data from diverse acoustic conditions are available, then these approaches can be made robust against noise and reverberation. Following the recent success of deep learning based supervised learning methods in various signal processing related tasks \cite{Krizhevsky2012, Hinton2012}, different methods for DOA estimation have been proposed \cite{Ma2017, Vesperini2016, Takeda2016, Xiao2015, Takeda2016a, Chakrabarty2017a, Chakrabarty2017b}. A common aspect of the methods proposed in \cite{Ma2017, Vesperini2016, Takeda2016, Xiao2015, Takeda2016a} is that they all involve an explicit feature extraction step. In \cite{Xiao2015, Vesperini2016}, GCC vectors, computed from the microphone signals, are provided as input to the learning framework. In \cite{Takeda2016, Takeda2016a}, similar to the computations involved in the MUSIC method for localization, the eigenvalue decomposition of the spatial correlation matrix is performed to get the eigenvectors corresponding to the noise subspace, and is provided as input to a neural network. In \cite{Ma2017}, a binaural setup is considered and binaural cues at different frequency sub-bands are computed and given as input. Such feature extraction steps generally lead to a high computational cost. Additionally, when features computed from the microphone signals are given as input the neural network mainly just learns the functional mapping from the features to the final DOA, which can possibly lead to a lack of robustness against adverse acoustic conditions. 

One of the main reasons for the success of deep learning has been the encapsulation of the feature extraction step into the learning framework. Also, by studying the traditional signal processing based methods for DOA estimation, it can be seen that most methods exploit the phase difference information between the microphone signals to perform localization. Based on these observations, in \cite{Chakrabarty2017a}, the current authors proposed a convolutional neural network (CNN) based supervised learning method for broadband DOA estimation of a single active speaker per short-time Fourier transform (STFT) time frame. Rather than involving an explicit feature extraction step, the phase component of STFT coefficients of the input signal were directly provided as input to the neural network. Another contribution of the work was to show the possibility of training the system using synthesized noise signals, which made the creation of training data much simpler compared to using real world signals like speech. 

Following that, in \cite{Chakrabarty2017b}, the previously proposed framework was extended to estimate multiple speaker DOAs. There, a novel method was developed to generate the training data using synthesized noise signals for multi-speaker localization. One of the main challenges of using noise signals for the multi-speaker case is that, for overlapping signals, the phase of the STFT coefficients get combined non-linearly, and depend on the magnitude of the individual signals. This makes the learning procedure for the CNN difficult. To overcome this problem, the property of W-disjoint orthogonality \cite{Rickard2002}, which holds approximately for speech signals, was utilized. In terms of evaluation, only preliminary results with simulated data for a single acoustic setup was shown in \cite{Chakrabarty2017b}.

In this paper, we further extend the initial work on DOA estimation of multiple speakers presented in \cite{Chakrabarty2017b}. The formulation of the task of multi-speaker DOA estimation as a multi-label multi-class classification problem is presented, where first the posterior probabilities of the active source DOAs are estimated at the frame-level. Then, these frame-level probabilities are averaged over multiple time frames depending on the chosen block length over which the final DOA estimates are to be obtained. From these averaged posterior probabilities, assuming the number of speakers, $L$, within that block is known, the DOAs corresponding to the classes with the $L$ highest probabilities are chosen as the final DOA estimates. To build robustness to adverse acoustic conditions, multi-condition training in the form of training data from diverse acoustic scenarios is performed. A detailed description of the previously proposed method for generating training data using synthesized noise signals is also presented. 

With respect to the proposed CNN architecture, we first posit that due to the small filters chosen to learn the phase correlations between neighboring microphones, $M-1$ convolution layers are required to learn from the phase correlation between all the microphone pairs, where $M$ is the number of microphones in the array. Through experimental evaluation, the requirement of $M-1$ layers is shown in terms of both localization performance as well as number of trainable parameters. The influence of distance between the sources and the microphone array is also investigated experimentally. Through further experiments with both simulated and measured room impulse responses (RIRs), the robustness of the proposed method to unseen acoustic conditions and noise types is investigated. Additionally, we also show that even when the CNN is trained to estimate the posterior probabilities of maximum two DOA classes per STFT time frame, at a block level the proposed method can be used to localize greater than two speakers also. 

The remainder of this paper is organized as follows. In Section~\ref{sec:Prob} the formulation of the problem as a multi-class multi-label classification is described. In Section~\ref{sec:Inrep}, we review the input feature representation used in our framework. The task of obtaining the final DOA estimates in our proposed system is described in Section~\ref{sec:CNN}. Section~\ref{sec:Noise} presents a detailed description of the proposed method for generating training data using synthesized noise signals. Experimental evaluation of the proposed method is presented in Section~\ref{sec:Exp}. Section~\ref{sec:con} concludes the paper.    

\section{Problem Formulation}
\label{sec:Prob}

We want to utilize a CNN based supervised learning framework for estimating the DOAs of multiple simultaneously active sources by learning the mapping from the recorded microphone signals to the DOA of the active speech sources using a large set of labeled data. The DOA estimation in this work is performed for signal blocks that consist of multiple time frames of the STFT representation of the observed signals. The block length can be chosen depending on the application scenario. For example, for dynamic sound scenes it might be preferable to choose shorter block lengths compared to a scenario when it is known that the sources would be static. 

The problem of multi-source DOA estimation is formulated as an $I$-class multi-label classification problem. As the first step, the whole DOA range is discretized to form a set of possible DOA values, $\Theta = \lbrace \theta_{1}, \ldots, \theta_{I} \rbrace$. A class vector of length $I$ is then formed where each class corresponds to a possible DOA value in the set $\Theta$. In this work, we assume an independent source DOA model, i.e., the spatial location of the sources are independent of each other. Due to this assumption, multi-label classification can be tackled using the binary relevance method \cite{Read2011}, where the assignment of each DOA class label to the input is treated as a separate binary classification problem. As stated earlier, the aim is to obtain the DOA estimates of multiple speakers for a signal block, however the input to the system is a feature representation for each STFT time frame separately.   

As shown in Fig.~\ref{fig:BD}, a supervised learning framework consists of a training and a test phase. In the training phase, the CNN is trained with a training data set that consists of pairs of fixed dimension feature vectors for each STFT time frame and the corresponding true DOA class labels. In the test phase, given the input feature representation corresponding to a single STFT time frame, the first task is to estimate the posterior probability of each DOA class. Following this, depending on the chosen block length, the frame-level probabilities are averaged over all the time frames in the block. Finally, considering $L$ sources, the DOA estimates are given by selecting the $L$ DOA classes with the highest probabilities. 

In this work, we consider the number of sources $L$ to be known. As an alternative, the number of active sources can be estimated based on the number of clear peaks in the averaged posterior probabilities for a signal block. Also, the recorded signal from a reference microphone can also be used for speaker count estimation using the method proposed in \cite{Stoeter2018}. Investigating the best strategy for this problem would be part of future work. 

\begin{figure}[t]
	\includegraphics[scale=0.45]{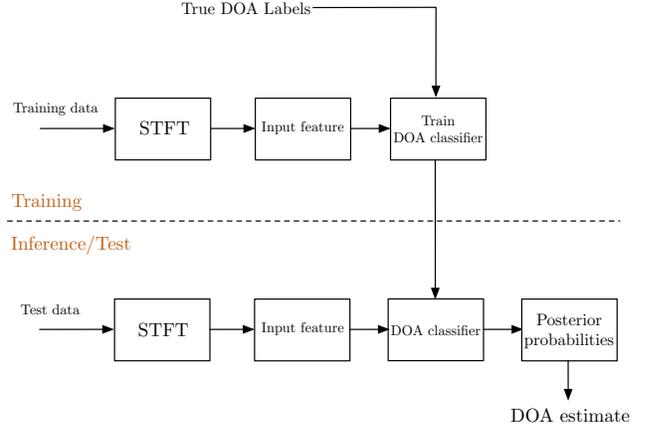}
	\caption{Block diagram of the proposed system.}
	\label{fig:BD}
\end{figure}
\section{Input Representation}
\label{sec:Inrep}
\begin{figure*}[t]
	\captionsetup{width=.99\textwidth}
	\centering
	\includegraphics[scale=0.53]{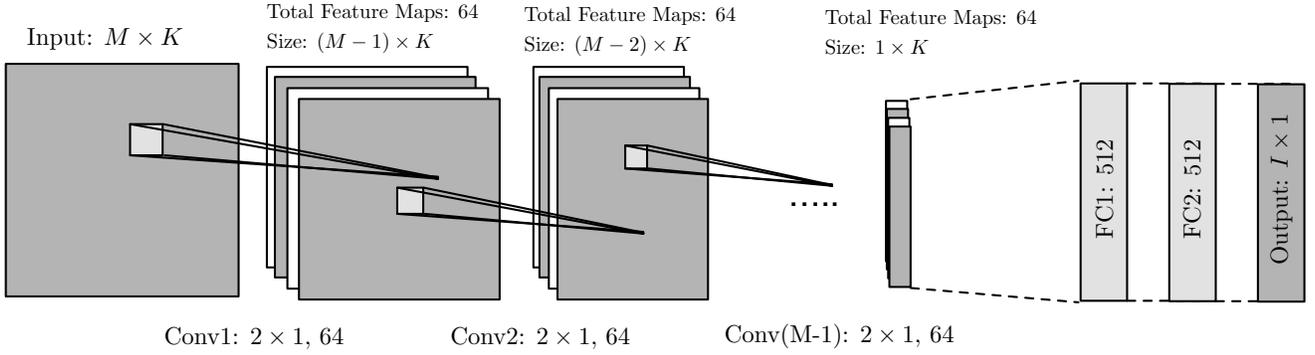}
	\caption{Proposed Architecture.}
	\label{fig:Arch}
\end{figure*}
In this work, the aim is to learn the relevant features for the task of DOA estimation via training rather than have an explicit feature extraction step to compute the input to be given to the system. Therefore we use the \emph{phase map} \cite{Chakrabarty2017a, Chakrabarty2017b} as the input feature representation in this work. For the sake of completeness, we give a brief description of this representation.

As described earlier, the input to the DNN framework is a feature representation corresponding to each STFT time frame. Let us consider that the received microphone signals are transformed to the STFT domain using an $N_f$ point discrete Fourier transform (DFT). In the STFT domain, the observed signals at each TF instance are represented by complex numbers. Therefore, the observed signal can be expressed as
\begin{equation}
Y_{m}(n,k) = A_{m}(n,k) e^{j\phi_{m}(n,k)} ,
\end{equation}
where $A_{m}(n,k)$ represents the magnitude component and $\phi_{m}(n,k)$ denotes the phase component of the STFT coefficient of the received signal at the $m$-th microphone for the $n$-th time frame and $k$-th frequency bin. In this work, we directly provide the phase component of the STFT coefficients of the received signals as input to our system. Note that this phase term consists of the phase of the source along with the effect of the propagation path. The idea is to make the system learn the relevant feature for DOA estimation from the phase component through training. 

Since the aim is to compute the posterior probabilities of the DOA classes at each time frame, the input feature for the $n$-th time frame is formed by arranging $\phi_{m}(n,k)$ for each time-frequency bin $(n,k)$ and each microphone $m$ into a matrix of size $M \times K$, where $K = N_{f}/2 +1$ is the total number of frequency bins, upto the Nyquist frequency, at each time frame and $M$ is the total number of microphones in the array. We call this feature representation as the \emph{phase map}. For example, if we consider a microphone array with $M=4$ microphones and $N_{f} = 256$, then the input feature matrix is of size $4 \times 129$. 

Given the input representations, the next task is to estimate the posterior probabilities of the $I$ DOA classes for each time frame. For this, we propose a CNN based supervised learning method, described in the following section.

\section{DOA estimation with CNNs}
\label{sec:CNN}

CNNs are a variant of the standard fully-connected neural network, where the architecture typically consists of one or more convolution layers followed by fully-connected networks leading ot the output \cite{LeCun1998}. In this work, the main motivation behind using CNNs is to learn the discriminative features for DOA estimation from the phase map input by applying small local filters to learn the phase correlations at the different frequency sub-bands. 

Given the phase map as the input, the CNN generates the posterior probability for each of the DOA classes. Let us denote the phase map for the $n$-th time frame as $\bm{\Phi}_{n}$. Then the posterior probability generated by the CNN at the output is given by $p(\theta_i|\bm{\Phi}_n)$, where $\theta_i$ is the DOA corresponding to the $i$-th class. In Fig.~\ref{fig:Arch}, the CNN architecture used in this work is shown. In the convolution layers, small filters of size $2 \times 1$ are applied to learn the phase correlations between neighboring microphones at each frequency sub-band separately. This is in contrast to \cite{Chakrabarty2017a}, where square filters of size $2 \times 2$ were used to learn the features from the neighboring frequency bins also. However, in the case of multiple speakers neighboring frequency bins might contain dominant activity from different speakers, therefore in this work we use $2 \times 1$ filters. These learned features for each sub-band are then aggregated by the fully connected layers for the classification task. The proposed architecture consists of at most $M-1$ convolution layers, where $M$ is the number of microphones, since after $M-1$ layers performing 2D convolutions is no longer possible as the feature maps become vectors. 

In terms of the design choice related to the number of convolution layers, we posit that by using small filters of size $2 \times 1$, with each subsequent convolution layer after the first one, for each sub-band, the phase correlation information from different microphone pairs are aggregated due to the growing receptive field of the filters, and to learn from the correlation between all microphone pairs, $M-1$ convolution layers would be required to incorporate this information into the learned features. In Section~\ref{sssec:convlayers}, we experimentally demonstrate that indeed $M-1$ convolution layers are required to obtain the best DOA estimation performance for a given microphone array and also show the efficiency of this design choice in terms of number of free parameters.   

As stated earlier, we utilize the binary relevance method \cite{Read2011} to tackle the multi-label classification problem, therefore the output layer of the CNN consists of $I$ sigmoid units, each corresponding to a DOA class. During training, the optimization of the network weights are done in terms of each output neuron separately, using binary cross-entropy as the loss function. 

Here, the task of multi-source DOA estimation is performed for a signal block consisting of $N$ time frames. The block-level posterior probability is obtained by averaging $N$ frame-level posterior probabilities for each $\theta_{i}$, given by
\begin{equation} \label{eq:avg}
p_{n}(\theta_{i}) = \frac{1}{N} \sum_{n}^{n+N-1} p(\theta_i|\bm{\Phi}_n).
\end{equation} 
From these computed average posterior probabilities, the $L$ DOAs corresponding to the $L$ classes with the highest probabilities are selected as the DOA estimates. In this work we chose this simple method to demonstrate the effectiveness of the proposed algorithm. Using more advanced post-processing methods, such as automatic peak detection \cite{Cormen2009}, is beyond the scope of this paper. 

\section{Training Data Generation}
\label{sec:Noise}
 
In this section, we describe the training data generation method employed in this work. Please recall that though the task of DOA estimation is performed for a segment of multiple time frames, in the proposed system the posterior probabilities of the DOA classes are estimated at each time frame. Therefore, using speech as training signals can be problematic since we would require an extremely accurate voice activity detection method in order to avoid including silent time frames in the training data, and errors in this task can adversely affect the training. To avoid this problem, in \cite{Chakrabarty2017a}, we proposed to use synthesized noise signals to generate the training data for the single speaker scenario. However, when trying to localize simultaneously active speakers, using overlapping noise signals for the training data is not suitable since at each TF bin, the phase component of the observed microphone signals' STFT coefficient is a non-linear combination of the phase of the individual directional sources. Thus, learning the relevant features from such an input might be difficult for the CNN. 

To effectively use synthesized noise signals to generate the training data, and taking into account the aim to localize speech sources, we utilize the assumption that the TF representation of two simultaneously active speech sources do not overlap. This is known as W-disjoint orthogonality, and, with an appropriate choice of the time and frequency resolutions, has been shown to hold approximately for speech signals \cite{Rickard2002}. In the following, we explain the procedure for generating the training data for a scenario with two active speakers.

As a first step, we generate the training signals for a single speaker case by convolving the room impulse responses (RIRs) corresponding to different directions for each acoustic condition considered for training with synthesized spectrally white noise signals. Then, for a specific source array setup, the STFT representation of two multi-channel training signals, corresponding to different DOAs, are concatenated along the time frame axis. Following this, for each frequency sub-band separately, the time-frequency bins for all microphones are randomized to get a single training signal. This procedure is repeated for all combinations of DOAs for all different acoustic conditions considered for training. Finally, the phase map corresponding to each time frame, for all training signals, is extracted to form the complete training dataset. 

While generating the training data, there are two important things to note regarding the randomization process. First, it is essential that the randomization of the TF bins is done separately for each frequency sub-band, such that the order of the frequency sub-bands remains the same for different time frames. This is essential since phase correlations are frequency dependent and for all the different time frames, preserving the spectral structure can aid the feature learning. Secondly, it is essential that for each frequency sub-band, the TF bins for all the microphones are randomized together, such that phase relations between the microphones for the individual TF bins are preserved.   

An illustration of this procedure is shown in Fig.~\ref{fig:Tr_illustration}. The figure on the left illustrates the concatenated TF representation of two directional signals, originating from two different directions, $\theta_1$ and $\theta_2$. Following the randomization procedure, it can be seen that at each time frame there are approximately equal number of TF bins with activity corresponding to the two DOAs. Therefore, at each frequency sub-band of the phase map input to the CNN, the phase of the STFT coefficients for all microphones correspond to a single source. This makes the assumption of disjoint activity of signals implicit within our framework. With this training input, the CNN can learn the relevant features for localizing multiple speakers at each time frame from the individual TF bins that contain the phase relations across the microphones for each source DOA separately.  

By repeating the above mentioned procedure for all possible angular combinations and acoustic conditions, we obtain the complete training dataset. The different acoustic conditions considered for the multi-condition training of the CNN is given in Table.~\ref{tab:Train}. The different rooms as well as positions inside each room are considered to develop robustness in various acoustic conditions, whereas additionally the network is also trained with different levels of spatially white noise for robust performance in noisy scenarios. 

In total, the training data consisted of around 12.4 million time frames. The CNN was trained using the Adam  gradient-based optimizer \cite{Kingma2014}, with mini-batches of 512 time frames and a learning rate of 0.001. During training, at the end of the convolution layers and after each fully connected layer, a dropout procedure \cite{Srivastava2014} with a rate of 0.5 was used to avoid over fitting. All the implementations were done in Keras \cite{chollet2015keras}. 

Please note that, in this work, the CNN is trained to estimate the posterior probabilities of DOAs of only two speakers given the phase map input for each STFT time frame. By following the same procedure as described above the method can be extended for estimating the DOA class posterior probabilities of more than two speakers per time frame. In Section \ref{sssec:dyn}, it is shown that despite such a training procedure the proposed method can estimate the DOAs of more than two speakers for a signal block with multiple time frames. 
\begin{table*}[t]
	\captionsetup{width=.99\textwidth}
	\normalsize
	\centering
	\setlength{\tabcolsep}{8 mm}
	\vspace{3 em}
	\caption{Configuration for training data generation. All rooms are 2.7 m high.}
	\begin{tabular}{c|c}
		\hline \hline
		\multicolumn{2}{c}{{Simulated training data}}	\\
		\hline 
		Signal & Synthesized noise signals \\
		\hline
		Room size  & R1: ($6 \times 6$) m , R2: ($5 \times 4$) m, R3: ($10 \times 6$) m, R4: ($8 \times 3$) m, R5: ($8 \times 5$) m \\
		\hline
		Array positions in room & 7 different positions in each room \\
		\hline
		Source-array distance& 1 m and 2 m  for each array position \\
		\hline
		RT$_{60}$ (s) & R1: 0.3, R2: 0.2, R3: 0.8, R4: 0.4, R5: 0.6 \\
		\hline
		SNR & Uniformly sampled from 0 to 30 dB \\		
		\hline \hline
	\end{tabular}
	\vspace{1 em}
	\label{tab:Train}\vspace{0.1em}
\end{table*}
%

\begin{figure}[t]
	\centering
	\includegraphics[scale=0.25]{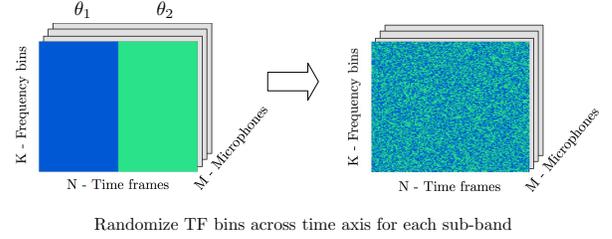}
	\caption{Illustration of the method used for generating the training data.}
	\label{fig:Tr_illustration}
\end{figure}
%

\begin{table}[t]
	\footnotesize
	\centering
	\setlength{\tabcolsep}{3 mm}
	\vspace{1 em}
	\caption{Configuration for generating test data for experiments presented in Section \ref{sssec:unseen} and \ref{sssec:noiseunseen}. All rooms are 3 m high.} 
	\begin{tabular}{c|c }
		\hline \hline
		\multicolumn{2}{c}{{Simulated test data}}	\\
		\hline 
		Signal & Speech signals from LIBRI \\
		\hline
		Room size  & Room 1: ($5 \times 7$) m , Room 2: ($9 \times 4$) m \\
		\hline
		Array positions in room & 4 arbitrary positions in each room \\
		\hline
		Source-array distance& 1.3 m  for Room 1, 1.7 m for Room 2\\		  
		\hline
		RT$_{60} (s)$ & Room 1: 0.38 , Room 2: 0.70 \\	
		\hline \hline
	\end{tabular}
	\label{tab:Test}\vspace{0.1em}
\end{table}
%
%
\begin{table*}[th]
	\captionsetup{width=.99\textwidth}
	\small
	\caption{Results for two different rooms with varying levels of spatially white noise computed over 3150 speech segments of 0.8 s for each array position. For each SNR, the result is averaged over the four different array positions in the room. }
	\setlength{\tabcolsep}{3 mm}
	\begin{center}
		\begin{tabular}{l|cc|cc|cc|cc|cc|cc}
			\midrule
			Test Room & \multicolumn{6}{c}{Room 1} & \multicolumn{6}{c}{Room 2}  \\
			\cmidrule(r){2-7} \cmidrule(r){8-13} 
			SNR & \multicolumn{2}{c}{$10$ dB} & \multicolumn{2}{c}{$20$ dB} & \multicolumn{2}{c}{$30$ dB} & \multicolumn{2}{c}{$10$ dB} & \multicolumn{2}{c}{$20$ dB} & \multicolumn{2}{c}{$30$ dB} \\
			\cmidrule(r){2-3} \cmidrule(r){4-5} \cmidrule(r){6-7} \cmidrule(r){8-9} \cmidrule(r){10-11} \cmidrule(r){12-13}
			Measure & MAE & Acc.& MAE & Acc. & MAE & Acc. & MAE & Acc. & MAE & Acc. & MAE & Acc.\\
			\midrule
			SRP-PHAT    &  $26.7$ & $37.7$ & $22.7$ & $49.3$  & $18.9$ & $60.4$ &$29.2$ & $28.2$ & $27.6$ & $36.9$  & $21.6$ & $48.3$  \\
			MUSIC    &  $23.3$ & $43.4$ & $16.2$ & $63.4$  & $13.4$ & $71.3$ &$27.1$ & $37.7$ & $18.6$ & $51.7$  & $16.6$ & $59.6$  \\
			Proposed     &  $14.5$ & $73.5$ & $3.5$ & $93.2$ &  $1.5$ & $98.1$ & $16.8$ & $63.4$ & $4.3$ & $88.9$  & $2.7$ & $96.3$  \\
			\midrule
		\end{tabular}
	\end{center}
	\label{tab:simres}
\end{table*}

\section{Experimental Evaluation}
\label{sec:Exp}

In this section, different experiments with simulated and measured data are presented to objectively evaluate the performance of the proposed system. For all the experimental evaluations except the one presented in Section \ref{sssec:convlayers}, we consider a ULA with $M = 4$ microphones with inter-microphone distance of 8 cm, and the input signals are transformed to the STFT domain using a DFT length of $N_{f} = 512$, with $50\%$ overlap, resulting in $K = 257$. The sampling frequency of the signals is $F_{s} = 16$ kHz. To form the classes, we discretize the whole DOA range of a ULA with a $5^{\circ}$ resolution to get $I = 37$ DOA classes, for both training and testing. All the presented objective evaluations are for the two speakers scenario. However, in Section~\ref{sssec:dyn}, we also demonstrate the ability of the proposed method to deal with scenarios with varying number of speakers.

The speech signals used for evaluation are taken from the LIBRI speech corpus. With random selected speech utterances, five different two speaker mixtures, each of length 2 s, were used. Since the angular space is discretized with a 5$^{\circ}$ resolution, for the experiments with simulated RIRs in Section \ref{ssec:sim}, it was ensured that the angular distance between the two speakers in the different mixtures is at least 10$^{\circ}$. Therefore, for a specific source-array setup in a room, each two speaker mixture is considered for each possible angular combination. This was done to avoid influence of signal variation on the difference in performance for different acoustic conditions. 

Since the speech utterances can have different lengths of silence at the beginning, the central 0.8 s segment of the mixtures was selected for evaluation. Considering an STFT window length of 32 ms with 50$\%$ overlap, this resulted in a signal block of $N= 50$ time frames over which the frame-level posterior probabilities are averaged for the final DOA estimation, as shown in (\ref{eq:avg}).

\begin{table*}[th]
	\captionsetup{width=.99\textwidth}
	\small
	\caption{Results for two different rooms with varying levels of babble noise computed over 3150 speech segments of 0.8 s for each array position. For each SNR, the result is averaged over the four different array positions in the room.}
	\setlength{\tabcolsep}{3 mm}
	\begin{center}
		\begin{tabular}{l|cc|cc|cc|cc|cc|cc}
			\toprule
			Test Room & \multicolumn{6}{c}{Room 1} & \multicolumn{6}{c}{Room 2}  \\
			\cmidrule(r){2-7} \cmidrule(r){8-13} 
			SNR & \multicolumn{2}{c}{$-5$ dB} & \multicolumn{2}{c}{$0$ dB} & \multicolumn{2}{c}{$5$ dB} & \multicolumn{2}{c}{$-5$ dB} & \multicolumn{2}{c}{$0$ dB} & \multicolumn{2}{c}{$5$ dB} \\
			\cmidrule(r){2-3} \cmidrule(r){4-5} \cmidrule(r){6-7} \cmidrule(r){8-9} \cmidrule(r){10-11} \cmidrule(r){12-13}
			Measure & MAE & Acc.& MAE & Acc. & MAE & Acc. & MAE & Acc. & MAE & Acc. & MAE & Acc.\\
			\midrule
			SRP-PHAT    &  $22.4$ & $40.8$ & $21.8$ & $46.1$  & $19.9$ & $57.8$ &$23.7$ & $40.2$ & $20.8$ & $46.6$  & $20.1$ & $48.3$  \\
			MUSIC    &  $23.9$ & $39.2$ & $18.8$ & $49.4$  & $16.3$ & $59.9$ &$25.9$ & $36.3$ & $19.2$ & $49.9$  & $18.1$ & $52.1$  \\
			Proposed     &  $5.0$ & $91.9$ & $2.1$ & $96.8$ &  $1.1$ & $98.7$ & $7.1$ & $82.9$ & $3.4$ & $94.3$  & $2.0$ & $97.5$  \\
			\bottomrule
		\end{tabular}
	\end{center}
	\label{tab:babres}
\end{table*}

\subsection{Baselines and objective measures}
\label{ssec:base}
The performance of the proposed method is compared to two commonly used signal processing based methods: Steered Response Power with PHase Transform (SRP-PHAT) \cite{Brandstein1997}, and broadband MUltiple SIgnal Classification (MUSIC) \cite{Dmochowski2007}. For the broadband MUSIC method, to keep the comparison similar with the other methods, the MUSIC pseudo-spectrum is computed at each frequency sub-band for each STFT time frame, with an angular resolution of 5$^{\circ}$ over the whole DOA space, and then it is averaged over all the frequency sub-bands to get the broadband pseudo-spectrum. This is then averaged over all the time frames considered in a signal block and similar to the proposed method, the $L$ DOAs with the highest values are selected as the final DOA estimates. Similar post-processing is also performed for the computed SRP-PHAT pseudo-likelihoods at each time frame to get the final DOA estimates for a signal block.

For the objective evaluation, two different measures were used: Mean Absolute Error (MAE) and localization accuracy (Acc.). The mean absolute error computed between the true and estimated DOAs for each evaluated acoustic condition is given by 
\begin{equation}
\text{MAE}(^\circ) = \frac{1}{LC} \; \sum_{c =1}^{C} \, \sum_{l=1}^{L} | \theta_{l}^{c} - \widehat{\theta}_{l}^{c} |,
\end{equation}
where $L$ is the number of simultaneously active speakers and $C$ is the total number of speech mixture segments considered for evaluation for a specific acoustic condition. The true and estimated DOAs for the $l$-th speaker in the $c$-th mixture are denoted by $\theta_{l}^{c}$ and $\widehat{\theta}_{l}^{c}$, respectively. 

The localization accuracy is given by
\begin{equation}
\text{Acc.}(\%) = \frac{\widehat{C}^{}_{\text{acc.}}}{C} \times 100,
\end{equation}
where $\widehat{C}^{}_{\text{acc.}}$ denotes the number of speech mixtures for which the localization of the speakers is accurate. In our evaluation, the localization of speakers for a speech segment is considered accurate if the distance between the estimated and the true DOA for all the speakers is less than or equal to 5$^{\circ}$. 
\subsection{Experiments with simulated RIRs}
\label{ssec:sim}

In this section, first, the performance of the proposed method is evaluated for acoustic conditions different from those considered during training, in the presence of varying levels of spatially uncorrelated white noise in Section \ref{sssec:unseen}. Then, we evaluate the performance in the presence of varying levels of diffuse babble noise, a noise type which was unseen during training, along with a constant level of spatially white noise in Section \ref{sssec:noiseunseen}. In Section \ref{sssec:convlayers}, we study the influence of the number of convolution layers on the performance of the proposed method and empirically demonstrate the optimal choice for the number of convolution layers for the proposed method.   

\begin{figure}[t]
	\centering
	\includegraphics[scale=0.45]{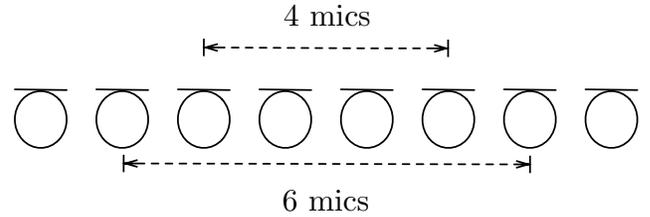}
	\caption{Array setup for experiment presented in Section \ref{sssec:convlayers}.}
	\label{fig:exp3setup}
\end{figure}
\subsubsection{Generalization to unseen acoustic conditions}
\label{sssec:unseen}

To evaluate the performance of the methods for unseen acoustic conditions, we consider two rooms with different reverberation times as shown in Table \ref{tab:Test}. In each room, the ULA is placed at four different positions and for each of these array positions, the two speakers from each of the five considered mixtures are placed at different angular positions at the same specified source-array distance. For each array position, the total number of mixtures considered for evaluation is $C = 630*5 = 3150$, where $630$ corresponds to the number of possible angular combinations with the constraint of $10^{\circ}$ angular separation between the two speakers for each of the five mixtures. 

The performance of the three methods under test is evaluated for three different levels of spatially white noise, with input SNRs 10, 20 and 30 dB, for both the rooms and the results in terms of the two considered objective measures are presented in Table \ref{tab:simres}. The shown results for each input SNR was averaged over the four different array positions considered in each room. 

From the results, it can be seen that the proposed method is able to provide accurate localization performance in acoustic environments that were not part of the training data. For input SNR of 30 dB, it manages to localize both sources accurately in 98$\%$ of the speech mixtures and shows a very low MAE. As the noise level increases, the performance worsens, however it always provides a much better localization accuracy and much lower error compared to both MUSIC and SRP-PHAT.

Considering same noise level, performance of the proposed method in both rooms is relatively similar compared to the signal processing based methods which have a considerably better performance in the less reverberant room (Room 1). One of the main reasons for this difference is the assumption of free-field sound propagation in the formulation of the signal processing based methods which leads to considerable deterioration in their performance in more reverberant conditions. On the other hand, the proposed supervised learning based method is trained in a diverse set of acoustic conditions, leading to a much better robustness to adverse acoustic environments. 

Overall, it can be seen that the proposed method has a superior performance, in terms of both MAE and localization accuracy, compared to the traditional signal processing based methods for all the different levels of spatially white noise in both rooms. Among the two signal processing based methods, MUSIC performs much better since the averaged broadband MUSIC pseudo-spectrum contains clearer peaks compared to SRP-PHAT which tends to exhibit a flatter distribution over the DOAs.   


\begin{figure*}[th]
	\captionsetup{width=.99\textwidth}
	\centering
	\subfigure[MAE]{\label{subfig:maeexp3}\includegraphics[width=0.38\textwidth]{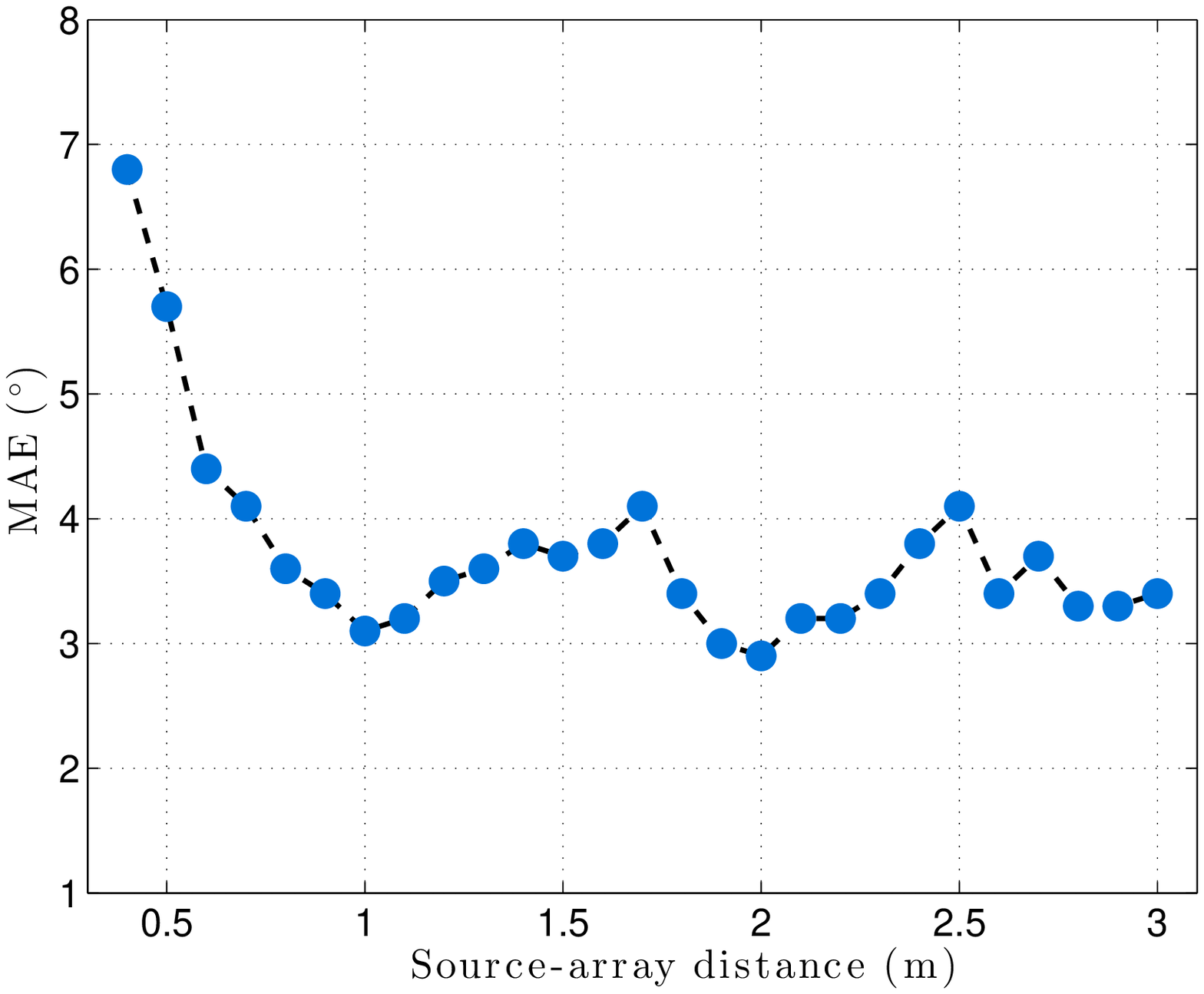}}
	\hspace{3em}
	\subfigure[Acc.]{\label{subfig:accexp3}\includegraphics[width=0.38\textwidth]{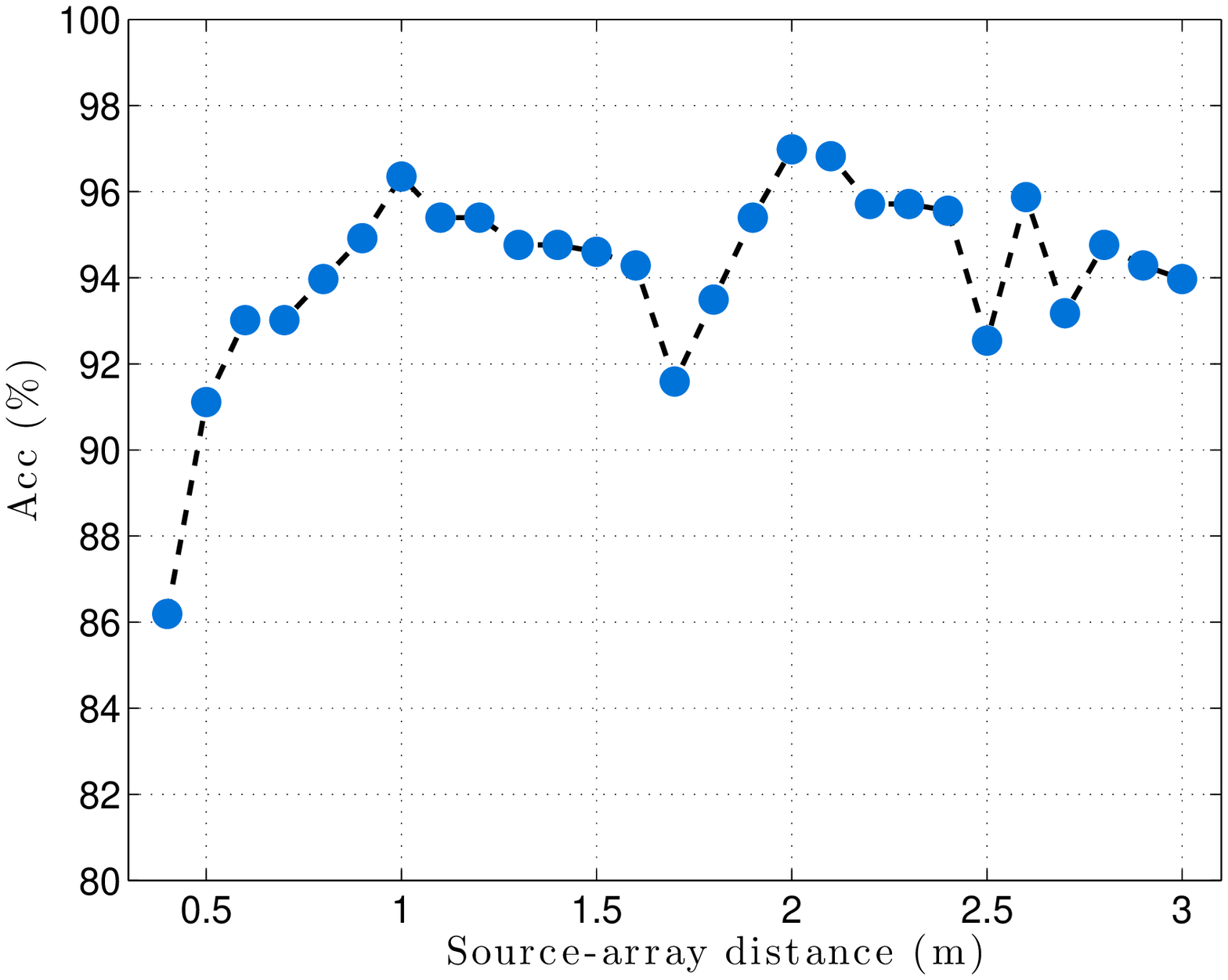}}
	%
	\caption{Results for the experiment showing the performance of the proposed method for increasing source-array distances presented in Section \ref{sssec:dist}.}
	\label{fig:Exp_dist}
\end{figure*}


\begin{figure*}[th]
	\captionsetup{width=.99\textwidth}
	\centering
	\subfigure[MAE]{\label{subfig:maeexp3}\includegraphics[width=0.4\textwidth]{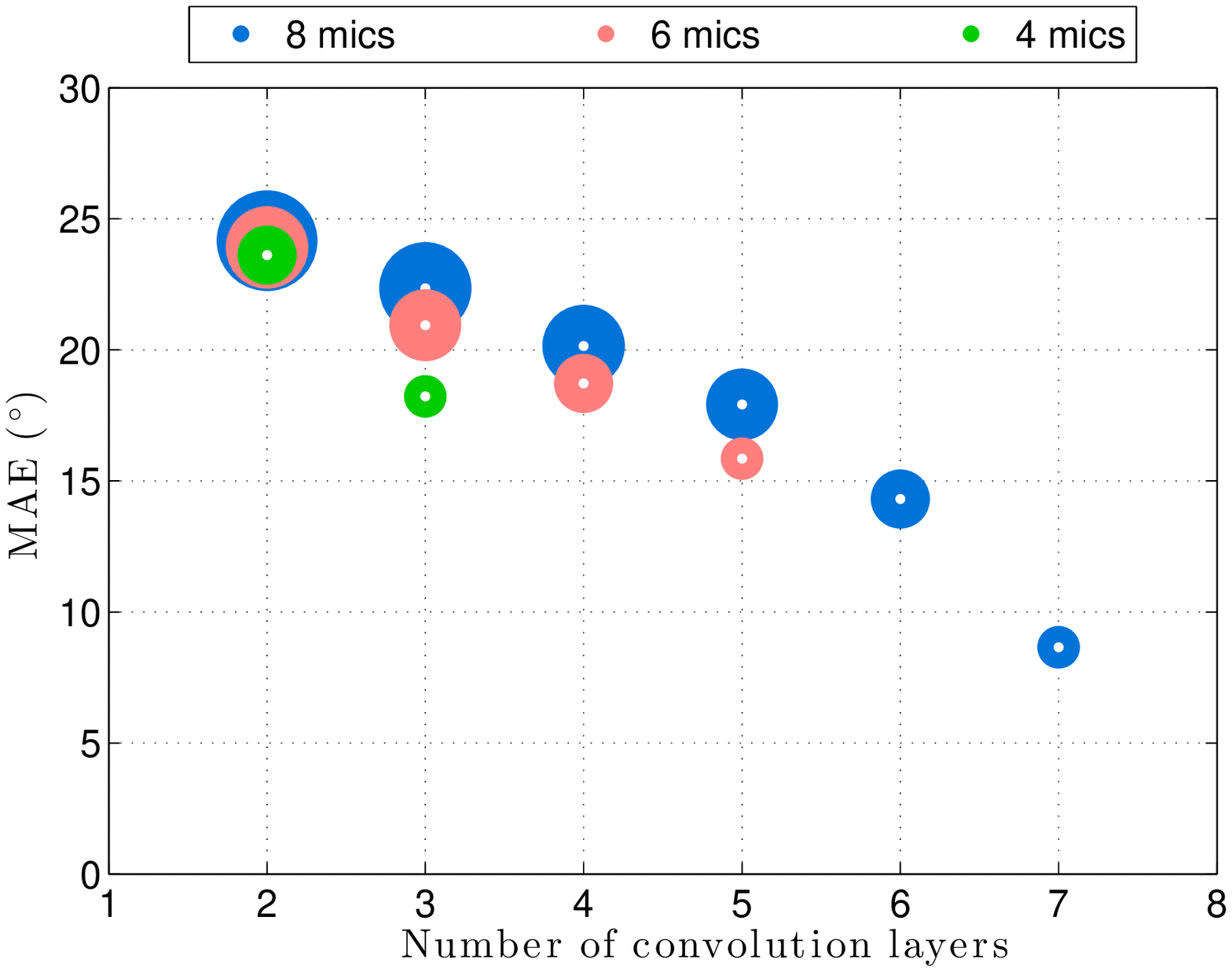}}
	\hspace{3em}
	\subfigure[Acc.]{\label{subfig:accexp3}\includegraphics[width=0.4\textwidth]{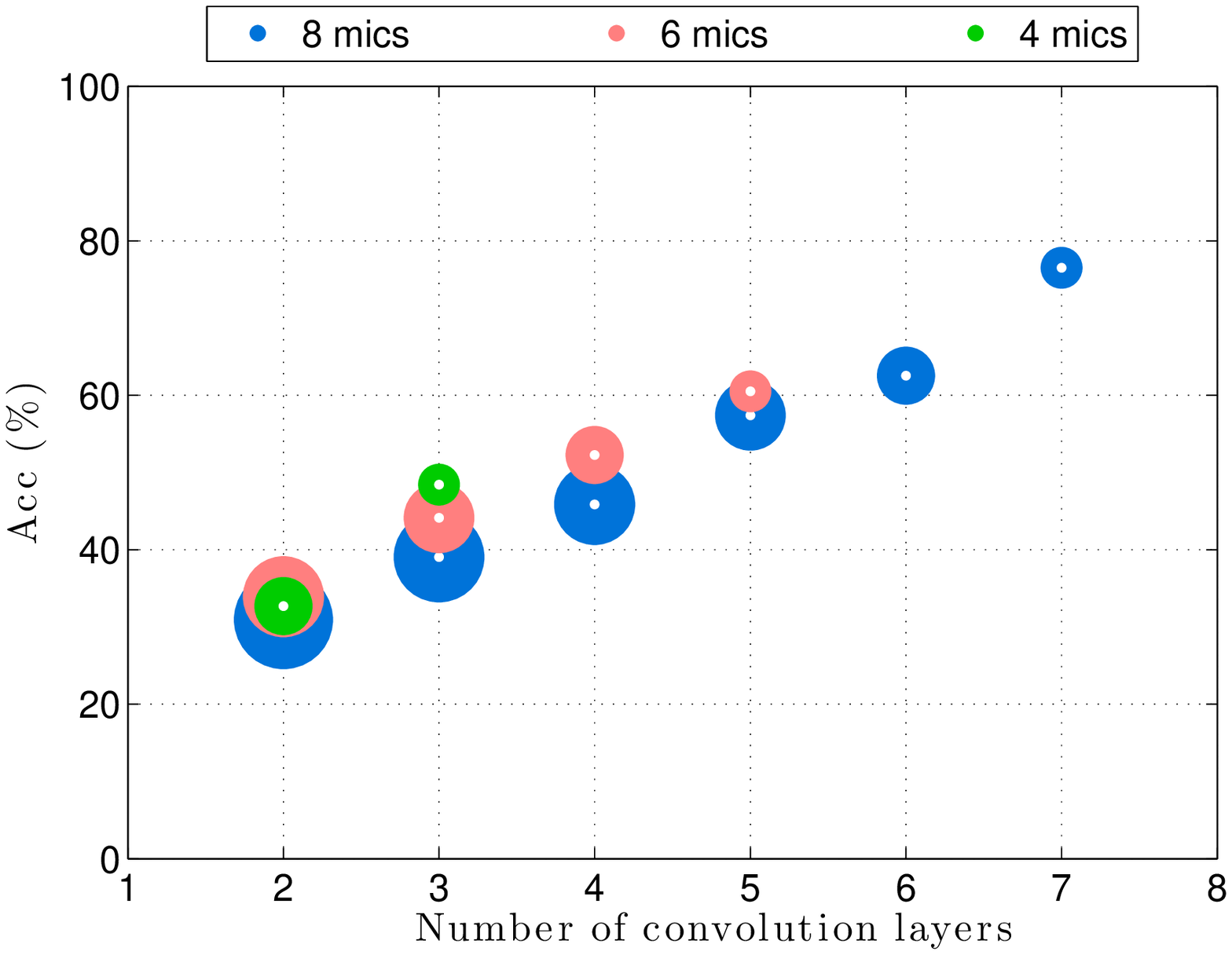}}
	%
	\caption{Results for the experiment on the influence of convolution layers on the proposed method presented in Section \ref{sssec:convlayers}.}
	\label{fig:Exp3}
\end{figure*}

\subsubsection{Generalization to unseen noise type}
\label{sssec:noiseunseen}

In the previous experiment, the performance of the proposed method was evaluated for different levels of spatially white noise, which is a noise type seen by the network during training. In this Section, we consider the presence of diffuse babble noise in the acoustic environment, which has different spatial as well as spectral characteristics compared to white noise, and is a noise type with which the CNN was not trained. A 40 s long sample of multi-channel diffuse babble noise was generated using the acoustic noise field generator \cite{Habets2007b}, assuming an isotropic spherically diffuse noise field. The generated babble noise was divided into 20 segments of 2 s each and randomly chosen segments were added to each mixture. 

The performance of the methods was evaluated for three different input SNRs of babble noise: -5 dB, 0 dB and 5 dB. Along with diffuse babble noise, spatially white noise with an input SNR of 40 dB was also added and results for the two different rooms are shown in Table \ref{tab:babres}. Similar to previous experiment, results for each input SNR of babble noise was averaged over the four different array positions considered in each room.  

Though the proposed method is not trained with diffuse babble noise, it can be seen from the results that even at the lowest input SNR of -5 dB, the proposed method is able to perform accurate localization of the two speakers in both rooms for approximately 90$\%$ of the speech mixtures. Since we consider an isotropic spherically diffuse noise field, the spatial coherence of the babble noise is frequency dependent whereas white noise is incoherent for all frequencies. Despite this difference, since the proposed method is trained to localize directional sources and due to multi-condition training, as long as the noise source is not directional the proposed method can provide very good performance.

If the results from Table \ref{tab:simres} are compared to Table \ref{tab:babres}, it can be seen that the deterioration in performance of the proposed method, in terms of the objective measures, as the noise levels increase is more prominent when white noise is considered compared to diffuse babble noise. The main reason for this difference is the spectral characteristics of the two different types of noises. On one hand, spatially white noise is present across the spectrum, therefore the input features at all frequency sub-bands are equally affected. On the other hand, babble noise is mostly dominant at low frequencies, therefore since each filter kernel in the convolution layers of the CNN learns from the complete input feature space, the filters are able to extract the relevant features for localization from the high SNR regions of the input to compensate for the lack of information in the low SNR regions. 

Overall, the proposed method provides a much better localization accuracy and lower error than the signal processing based methods, with the difference in performance being especially significant at low input SNRs of diffuse babble noise. 

\subsubsection{Influence of source-array distance}
\label{sssec:dist}

The CNN used for the earlier evaluations was trained for each room and array position for two specific source-array distances of 1 m and 2 m. To investigate the influence of source-array distance, in this part, the localization performance of the proposed method is evaluated for varying source-array distances.  

For this experiment, we simulated a room with dimensions $10 \times 11 \times 3$ m$^{3}$ and a reverberation time of 0.38 s. The test data was generated for three different array positions. For each of these array positions, the sound sources were placed at distances varying from 0.4 m to 3 m. It should be noted that both the speakers were placed at the same distance for each setup. A single two speaker mixture was used and spatially white noise was added resulting in input SNR of 20 dB. 

The results for this experiment, in terms of both MAE and localization accuracy, is shown in Fig. \ref{fig:Exp_dist}. Each point in the plot corresponds to a specific source-array distance. For each of these points, the measures were averaged over all possible angular combinations for the two speakers at each of the different array positions in the room. 

From the result plots, it can be seen that when the sources are very close to the microphone array the error in localization is higher, since the CNN was trained considering a far-field scenario, however for very small source-array distances, the sources are essentially in the near-field of the array. The minimum error as well as maximum accuracy in localization can be observed for the two specific distances of 1 m and 2 m, which were part of the training setup. Additionally, for distances close to these training distances, the errors are also relatively lower. When the sources are between the two training distances, the errors are slightly higher, however if we observe the absolute value of the MAE as well as the accuracy, the degradation in performance is not significant. Similarly for distances larger than 2 m, it can be seen that the localization performance deteriorates slightly. 

Overall, observing the absolute value of the objective measures, it can be seen that though the network is trained with two specific source-array distances, there is small deterioration in performance for other distances, except when the sources are very close to the microphone array.   
\begin{table*}[t]
	\captionsetup{width=.99\textwidth}
	\small
	\caption{Results with measured RIRs.}
	\setlength{\tabcolsep}{3 mm}
	\begin{center}
		\begin{tabular}{l|cc|cc|cc|cc|cc|cc}
			\toprule
			RT$_{60}$ & \multicolumn{4}{c}{0.160 s} & \multicolumn{4}{c}{0.360 s} & \multicolumn{4}{c}{0.610 s} \\
			\cmidrule(r){2-5} \cmidrule(r){6-9} \cmidrule{10-13} 
			Distances & \multicolumn{2}{c}{1 m} & \multicolumn{2}{c}{2 m} & \multicolumn{2}{c}{1 m} & \multicolumn{2}{c}{2 m} & \multicolumn{2}{c}{1 m} & \multicolumn{2}{c}{2 m} \\
			\cmidrule(r){2-3} \cmidrule(r){4-5} \cmidrule(r){6-7} \cmidrule(r){8-9} \cmidrule(r){10-11} \cmidrule(r){12-13}
			Measure & MAE & Acc.& MAE & Acc. & MAE & Acc. & MAE & Acc. & MAE & Acc. & MAE & Acc.\\
			\midrule
			SRP-PHAT    &  $\mathbf{}12.8$ & $\mathbf{}75.0$ & $15.33$ & $64.2$  & $15.8$ & $61.8$ &$19.8$ & $49.2$ & $15.3$ & $57.4$  & $21.5$ & $42.9$  \\
			MUSIC    &  $\mathbf{}4.9$ & $\mathbf{}87.0$ & $9.33$ & $78.2$  & $10.4$ & $72.8$ &$15.2$ & $54.2$ & $11.3$ & $70.7$  & $18.5$ & $47.3$  \\
			Proposed     &  $1.9$ & $89.7$ & $3.4$ & $86.1$ &  $3.27$ & $88.2$ & $4.35$ & $79.9$ & $3.14$ & $85.5$  & $4.43$ & $80.2$  \\
			\bottomrule
		\end{tabular}
	\end{center}
	\label{tab:mesres}
\end{table*} 
    
\begin{figure*}[th]
	\captionsetup{width=.99\textwidth}
	\centering
	\subfigure[Frame level DOA probabilities for the proposed method (top) and MUSIC (middle). The ground truth DOAs and source activities for each segment are shown in the bottom figure.]{\label{subfig:frameexp5}\includegraphics[width=0.7\textwidth]{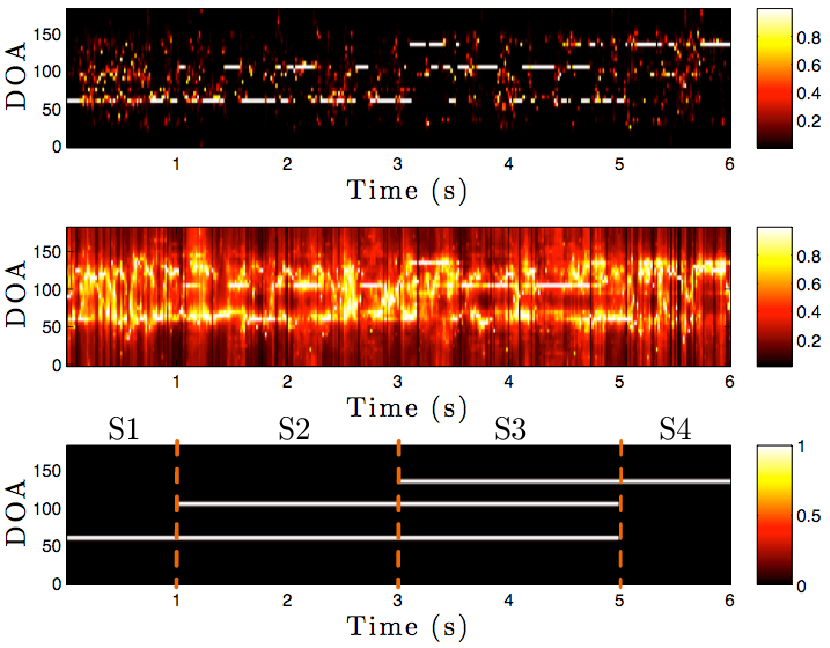}}
	\hspace{3em}
	\subfigure[Normalized histogram computed from the frame level probabilities for each segment.]{\label{subfig:avgexp5}\includegraphics[width=0.7\textwidth]{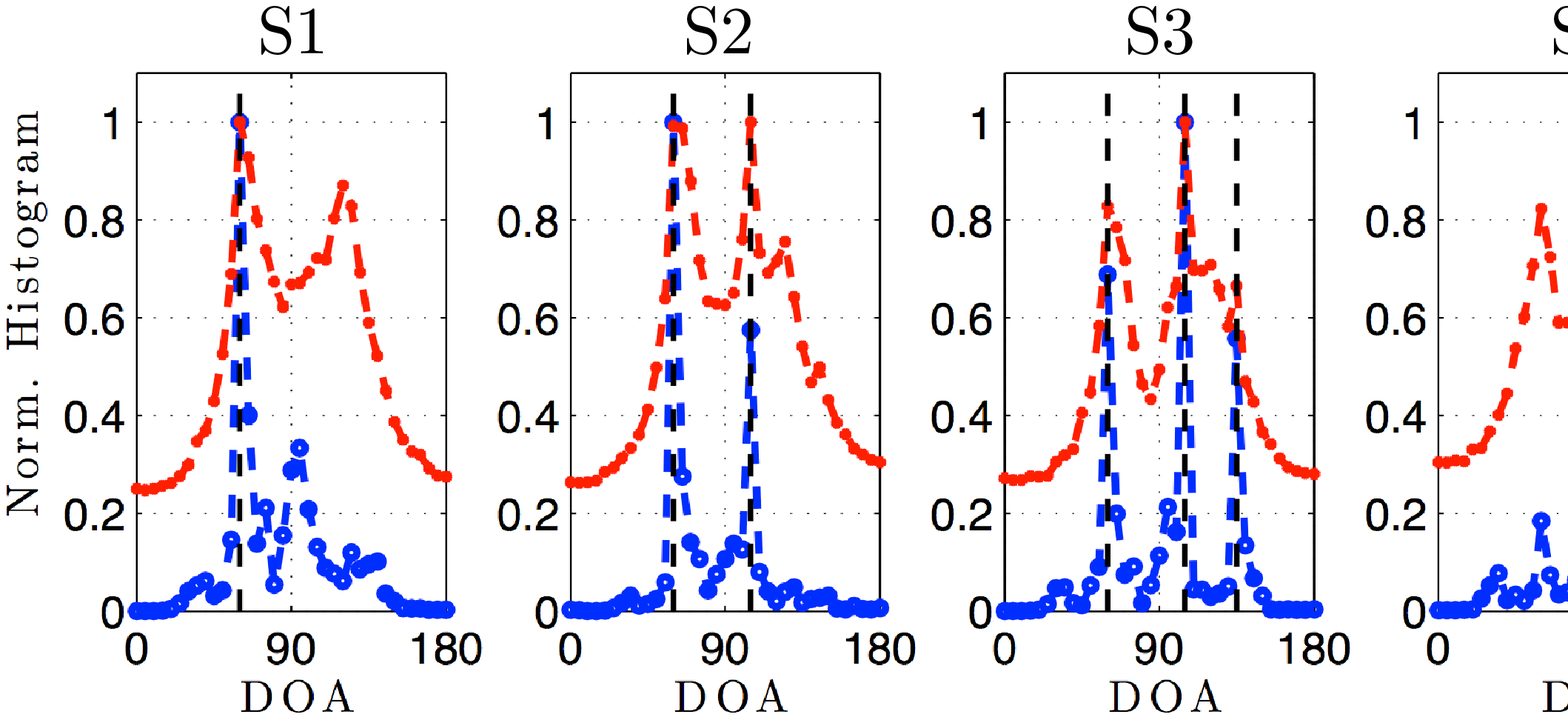}}
	%
	\caption{Results for experiment presented in Section \ref{sssec:dyn} with measured RIR and a four microphone ULA. The reverberation time of the room is 0.36 s with the source placed 2 m away from the array center. Spatially uncorrelated noise and diffuse babble noise were added to the mixture signal with input SNRs of 40 dB and 5 dB, respectively.}
	\label{fig:Exp5}
\end{figure*}

\subsubsection{Influence of number of convolution layers}
\label{sssec:convlayers}

In the previous experiments, we considered a ULA with $M =4$ microphones and the CNN architecture used was the same architecture that was proposed in \cite{Chakrabarty2017a,Chakrabarty2017b} which consisted of three convolution layers followed by two fully connected layers. In this section we empirically demonstrate that given the choice of small filters of size $2 \times 1$ for all the convolution layers, with the aim to learn the relevant features for localization from the phase correlations at neighboring microphones, a CNN architecture with three convolution layers is not always the best performing architecture. Here we show that the number of convolution layers need to be $M-1$ to obtain the best localization performance.  

For this experiment we consider a ULA with 8 microphones with an inter-microphone distance of 2 cm. From this array, we select two sub-arrays, one with 6 microphones and the other with 4 microphones that are formed by selecting the respective number of middle microphones from the main eight element array, as shown in Fig. \ref{fig:exp3setup}, to get a ULA with $M=6$ and another ULA with $M=4$, respectively. All the arrays have the same inter-microphone distance and array center. 

Using the same training data configuration from previous experiments (Table \ref{tab:Train}), multiple CNNs with number of convolution layers varying from 2 to $M-1$ are trained for each of the arrays. The number of convolution layers is restricted to $M-1$ since further 2D convolution layers are not possible as the microphone dimension of the phase map input is reduced to 1 after the $M-1$-th layer. For the eight microphone array, 6 CNNs are trained, whereas for the six microphones and the four microphone array, 4 and 2 CNNs are trained, respectively. All the networks were trained with the same amount of data. To analyze the performance of the 12 different trained networks, test data corresponding to the Room 1 configuration in Table \ref{tab:Test} is generated for each of the arrays. Spatially white noise is added for an input SNR of 30 dB. 

The results for this experiment, in terms of both MAE and localization accuracy, is shown in Fig. \ref{fig:Exp3}. In the figures, the center of the circle markers correspond to the value of the objective measure and the area of the markers denote the number of trainable/free parameters for that specific network. 

The first trend that can be noticed from the figures is that for each of the arrays, as the number of convolution layers is decreased from $M-1$ the performance of the networks degrades in terms of both MAE and localization accuracy. This shows that with small filters of size $2 \times 1$, to aggregate the phase correlation features from all the microphone pairs in an array, $M-1$ convolution layers are required. When lesser number of convolution layers are used, as the same filter size is used in each of these layers, phase correlation information from all microphone pairs are not incorporated into the learned features leading to deterioration in performance. 

It can also be seen from the figures that the best localization performances of the three arrays is different and it is better for the array with higher number of microphones. This difference in performance comes from the different apertures of the considered arrays, and similar to signal processing based localization methods, here also we observe better performance for a ULA with a larger aperture.  

In Fig. \ref{fig:Exp3}, we also observe that as the number of convolution layers is decreased the number of trainable/free parameters increases, as depicted by the area of the markers for each network. From Fig. \ref{fig:Arch}, it can be seen that when $M-1$ convolution layers are used, the size of each feature map at the end of the convolution layers is always $1 \times K$. As the number of convolution layers is decreased the size of each feature map at the end of the convolution layers actually becomes larger leading to a larger number of trainable/free parameters for the complete network. This further demonstrates the need of $M-1$ convolution layers, as very large number of free parameters can lead to problems of over fitting, if the amount of available training data is not sufficient.

Since the requirement of $M-1$ convolution layers is mainly related to the aggregation of information in the feature space by the slowly growing receptive field of the small filters used in our framework, techniques for a more aggressive expansion of the receptive field of the filters can also be employed. This is however beyond the scope of this paper and is a topic for future research.  

\subsection{Experiments with measured RIRs}
\label{ssec:meas}

For the experiments with measured RIRs, we used the Multichannel Impulse Response Database from Bar-Ilan University \cite{Hadad2014}. The database consists of RIRs measured at Bar-Ilan University's acoustics lab, of size $6 \times 6 \times 2.4$ m$^{3}$, for three different reverberation times of RT$_{60}$ = 0.160, 0.360, and 0.610 s. The recordings were done for several source positions placed on a spatial grid of semi-circular shape covering the whole angular range for a linear array, i.e., $[0^{\circ}, 180^{\circ}]$, in steps of 15$^{\circ}$ at distances of 1 m and 2 m from the center of the microphone array. 

The recordings were done with a linear microphone array with three different microphone spacings. For our experiment, we chose the [8, 8, 8, 8, 8, 8, 8] cm setup \cite{Hadad2014}, which consists of eight microphones where the distance between the microphones is 8 cm. We selected a sub-array of the four middle microphones out of the total eight microphones used in the original setup, to have a ULA with $M =4$ elements with an inter-microphone distance of $8$ cm, which corresponds to the array setup used in experiments with simulated RIRs. Therefore, the CNN trained with simulated data used for the earlier evaluations in Section~\ref{sssec:unseen} and \ref{sssec:noiseunseen} was also used for these experiments. We used the same five mixtures from earlier, with the total number of mixtures for evaluation being $C = 76*5 = 380$, where $76$ is the number of all possible angular combinations with discretization of the complete DOA space of a ULA with 15$^{\circ}$ resolution.        

The results for all the different reverberation times and source-array distances are shown in Table \ref{tab:mesres}. For this experiment, spatially white noise was added to each mixture resulting in an input SNR of 30 dB. 

Even when trained with simulated data only, the results show that the proposed method is able to provide very good localization performance in real conditions, even when the sources are placed far from the array in reverberant conditions. The performance of all the compared methods is better when the sources are close to the array, however the difference in performance, for different distances, for the signal processing based methods is considerable since the effect of reverberation is more significant when the sources are further away from the array. 

Overall, the proposed method provides significantly better performance compared to both MUSIC and SRP-PHAT, and the difference is more prominent as the acoustic environment becomes more reverberant.  

\subsubsection{Dynamic acoustic scenario}
\label{sssec:dyn}

In all the previous experiments, we considered the two speaker scenario for the evaluation of the performance of the proposed method. In this experiment we show that even though the CNN is trained to estimate the frame-level posterior probabilities of a maximum of two sources, with the proposed method it is possible to estimate the DOA of more than two sources for a short segment. Simultaneously, it is also shown that since the input to the CNN is the phase map for a single STFT time frame, the proposed method is also able to handle dynamic acoustic scenarios where the number of speakers changes over time.

For this experiment, we consider the reverberation time of 0.36 s and source-array distance of 2 m from the measured RIR database used in the previous experiment. A 6 s speech mixture segment is created where for the first 1 s only one source from 60$^{\circ}$ is active. For the next 2 s, an additional source is active from 105$^{\circ}$. A third source from 135$^{\circ}$ is active for the next 2 s along with the first two sources. For the final 1 s duration, only the third source is active. The source activities for each segment and the corresponding ground truth DOAs of the sources are shown in the bottom figure of Fig.~\ref{subfig:frameexp5}. Spatially white noise and diffuse babble noise are added to the speech mixture resulting in input SNRs of 40 dB and 5 dB, respectively. 

The estimated frame-level probabilities for the proposed method and broadband MUSIC are depicted in the top and middle figures of Fig.~\ref{subfig:frameexp5}, respectively. Since from the previous experiments, it was found that MUSIC is the better performing method out of the two considered signal processing based techniques, the results for SRP-PHAT are not presented. It can be seen that the estimated frame-level probabilities for the proposed method is much more concentrated towards the actual source DOAs compared to MUSIC. 

In Fig.~\ref{subfig:avgexp5}, the frame level probabilities are averaged over the time frames in each segment and then normalized to a maximum value of 1. This specific normalization is done for the purpose of visualization only. From these figures, it can be seen that the proposed method exhibits much clearer peaks at the true source DOAs compared to MUSIC which lead to the superior performance of the proposed method in previously presented evaluations even with the simple post-processing method considered in this work for obtaining the final DOA estimates. It can also be seen that in the segment S3, where three sources are simultaneously active, though the network is trained to estimate frame level probabilities of two speakers, clear peaks are visible at all the three true source DOAs. Also, when only one source is active (S1 and S4), the highest peaks correspond to the true DOA. 
   
\section{Conclusion}
\label{sec:con}

A convolutional neural network based supervised learning method for DOA estimation of multiple speakers was presented that is trained using synthesized noise signals. Through experimental evaluation, it was shown that the proposed method provides excellent localization performance in unseen acoustic environments as well as in the presence of unseen noise types. It was also shown to exhibit a far superior performance compared to the signal processing based localization methods, SRP-PHAT and MUSIC, for the tested conditions. The ability of the proposed method to deal with acoustic scenarios with varying number of sources was also shown. 

For the design choice of the number of convolution layers in the proposed architecture, it was empirically shown that for a microphone array with $M$ microphones, $M-1$ convolution layers are required for the best localization performance. It was also shown that such a choice leads to lesser number of trainable parameters. The choice of $M-1$ convolution layers is required for the aggregation of the phase correlation information from all microphone pairs in the extracted features, when using contiguous convolution operations, as done in this work.

\balance
\bibliographystyle{IEEEtran}
\bibliography{sapref_NIPS_taslp}
%
%
%
%
%
%
%
%
%

%
%

\end{sloppy}
\end{document}